\documentclass{jpp}
\usepackage{amsmath}
\usepackage{amssymb,amstext}
\usepackage{natbib}

\renewcommand{\vec}[1]{\boldsymbol{#1}}

\title[Nonlinear Alfv\'en-cyclotron waves]
  {On nonlinear Alfv\'en-cyclotron waves in multi-species plasma}

\author[E. Marsch and D. Verscharen]%
  {E\ls C\ls K\ls A\ls R\ls T\ns M\ls A\ls R\ls S\ls C\ls H
   \thanks{marsch@mps.mpg.de}\ns \and\ns
D\ls A\ls N\ls I\ls E\ls L\ns V\ls E\ls R\ls S\ls C\ls H\ls A\ls R\ls E\ls N
  \thanks{verscharen@mps.mpg.de}}

\affiliation{Max-Planck-Institut f\"ur Sonnensystemforschung, Max-Planck-Stra\ss{}e~2, D-37191~Katlenburg-Lindau, Germany}

\date{22 April 2010, revised 30 July 2010, accepted 17 August 2010}

\begin{document}

\maketitle

\begin{abstract}
Large-amplitude Alfv\'en waves are ubiquitous in space plasmas and a main
component of magnetohydrodynamic (MHD) turbulence in the heliosphere. As pump
waves, they are prone to parametric instability by which they can generate
cyclotron and acoustic daughter waves. Here, we revisit a related process
within the framework of the multi-fluid equations for a plasma consisting of
many species. The nonlinear coupling of the Alfv\'en wave to acoustic waves
is studied, and a set of compressive and coupled wave equations for the
transverse magnetic field and longitudinal electric field is derived for
waves propagating along the mean-field direction. It turns out that slightly
compressive Alfv\'en waves exert, through induced gyro-radius and
kinetic-energy modulations, an electromotive force on the particles in
association with a longitudinal electric field, which has a potential that is
given by the gradient of the transverse kinetic energy of the particles
gyrating about the mean field. This in turn drives electric fluctuations
(sound and ion-acoustic waves) along the mean magnetic field, which can
nonlinearly react back on the transverse magnetic field. Mutually coupled
Alfv\'en-cyclotron--acoustic waves are thus excited, a nonlinear process that
can drive a cascade of wave energy in the plasma, and may generate compressive
microturbulence. These driven electric fluctuations might have consequences
for the dissipation of MHD turbulence and, thus, for the heating and
acceleration of particles in the solar wind.
\end{abstract}

\section{Introduction}

Large-amplitude Alfv\'en waves are ubiquitous in space plasmas, and
particularly prominent in the solar wind \citep{tu95, bruno05}. They are
an essential component of magnetohydrodynamic (MHD) turbulence in the heliosphere
and known to originate mainly in the solar coronal holes \citep{cranmer09}.
As has been shown in the ample literature, an Alfv\'en mother (pump) wave is
prone to parametric instability \citep{stenflo76, derby78, goldstein78,
longtin86, brodin88, hollweg94, wong86, vinas91b, vinas91a,
stenflo00, ruderman04, stenflo07} by which it can generate cyclotron
and acoustic daughter waves that may undergo kinetic effects
\citep{araneda98} and collisionless Landau damping \citep{inhester90,
araneda07}. The continuous and wide interest in these waves also comes from
their astounding properties, namely that Alfve\'n-cyclotron waves, like
parallel magnetosonic-whistler waves, are nonlinear eigenmodes
\citep{sonnerup67, stenflo76} of the MHD, and
multi-fluid equations as shown below, for propagation along the mean magnetic
field.

Nonlinearly excited \citep{spangler89} acoustic waves appear to be common
in space plasmas as well, and density fluctuations \citep{tu95, bruno05}
are observed everywhere in the solar wind, although at a comparatively low
fluctuation level of merely a few percent. However, since compressive
fluctuations can be damped through kinetic effects, like Landau damping on
the thermal ions and electrons, they can provide an effective dissipation
mechanism for the nonlinear damping \citep{medvedev97} of Alfv\'en waves.
Consequently, the understanding of the coupling between Alfv\'enic wave
activity and density or charge-density fluctuations is of paramount interest
and importance in basic plasma physics, but alike in its applications to
nonlinear processes in space \citep{stenflo07} and astrophysical plasmas.

As will be shown in this paper, a coupled set of nonlinear second-order wave
equations for the transverse magnetic field, the transverse gyromotion of any
particle species in the multicomponent plasma considered, and the related
longitudinal electric field can be derived, which together describe the
wave--wave interactions and their mutual forcing. These equations provide a
physically and intuitively clear picture of the field and particle/plasma
dynamics and allow us to understand the results of recent hybrid simulations
of the parametric decay of Alfv\'en waves and their effects on the plasma
particles better. The main aim of this work is to provide algebraic
derivations and physical explanations. A numerical treatment of the full
equations to be derived subsequently appears promising, yet is beyond the
scope of this work.

In analytical  \citep{araneda07}, hybrid-simulation, and other numerical
simulation  \citep{araneda08} studies of the parametric instabilities of
Alfve\'n-cyclotron waves, it became obvious that ion trapping
 \citep{araneda08, araneda09} in the nonlinearly driven ion-acoustic waves
and pitch-angle scattering by the transverse daughter waves were found to
cause anisotropic heating of the proton core velocity distribution, and
simultaneously to create a proton beam along the mean field
 \citep{araneda08,valentini09}. These numerical results are in close
agreement with observed kinetic features in the solar wind and support the
observation that pitch-angle scattering  \citep{heuer07, marsch01} is the
key to understand the kinetic characteristics of thermal solar-wind protons.
But only recently convincing evidence has been found for ion-cyclotron waves
 \citep{jian09} to exist in the solar wind. Also simulations of electric
field spectra \citep{valentini08} have shown that the short-scale
termination of solar wind turbulence is characterized by the occurrence of
longitudinal electrostatic fluctuations. The spectra thus obtained seem to be
consistent with the electrostatic waves actually measured in the solar wind
 \citep{bale05} close to the Earth's bow shock, and in particular in the
ion-cyclotron range \citep{kellogg06}.

The present study will provide the foundation for insight into and further
study of the processes occurring at macroscopic and microscopic scales in
solar-wind turbulence, and thus will throw light on the related dissipation
processes through kinetic cascades and wave--particle interactions
\citep{marsch06}. The nonlinear equations derived here are used to describe
an elliptically polarized Alfv\'en wave as a simple but nontrivial example
of their application.

\section{The multi-fluid equations in conservation form}

\subsection{Fluid equations in the wave frame}

In this section, we shall first recapitulate the basic multi-fluid equations
\citep{goossens03} for a plasma consisting of electrons and various ionic
species. We start from the fundamental conservation laws, and then try to
make no approximations with respect to the field amplitudes in order to be
able to discuss and analyze nonlinear waves and convected wave-like
structures. We are going to use coordinates in the frame of reference moving
with the wave, which has a normal to its front denoted by $\hat{{\vec n}}$
and a propagation speed $\vec V = V \hat{{\vec n}}$ in the inertial frame, or
center of momentum frame that is defined below. This unit vector obeys the
relation $\hat{{\vec n}}^2 = 1$. The coordinate in this moving frame is
$\vec{\xi} = \vec{x} - \vec{V} t$, and all variables are assumed to depend on
space and time only through $\vec{\xi}$, and thus spatial and temporal
derivatives in the wave frame are reduced to derivatives with respect to
$\vec{\xi}$. Such a coordinate transformation has been used by many authors,
for example to study solitary waves in multi-ion plasmas
 \citep{hackenberg98}, or their stability properties  \citep{mckenzie93}.
Therefore, by using comoving coordinates, Maxwell's partial differential
equations in space and time, and similarly the fluid equations for the
different species, can be reduced to simpler differential equations in terms
of $\vec{\xi}$. The continuity equations thus reads
\begin{equation}
\frac{\partial}{\partial {\vec{\xi}}} \cdot \left( n_j \vec V_j \right) = 0,
\label{eq.2}
\end{equation}
with $\vec V_j$ being the flow velocity of species $j$ in the moving frame
$\vec V_j = \vec U_j - \vec V$, and $n_j$ is its number density.
With the charge denoted as $q_j$, we get the total charge density as
\begin{equation}
\sigma = \sum_j \sigma_j = \sum_j q_j n_j,
\label{eq.4}
\end{equation}
which must obey Gau\ss{}' law:
\begin{equation}
4\pi \sigma =  \frac{\partial}{\partial {\vec{\xi}}} \cdot \vec E.
\label{eq.5}
\end{equation}
Similarly, the total current density is given by
\begin{equation}
{\vec J} = \sum_j q_j n_j \vec U_j, \label{eq.6}
\end{equation}
which has to obey Amp\`{e}re's law:
\begin{equation}
{\vec J} = \frac{c}{4 \pi} \; \frac{\partial}{\partial {\vec{\xi}}}
\times \vec B + \frac{1}{4 \pi} \; \frac{\partial}{\partial
{\vec{\xi}}} \cdot ( \vec V \vec E ). \label{eq.7}
\end{equation}
The second term of Eq.~(\ref{eq.7}) is the displacement current. The conduction
minus convection current density may be written
\begin{equation}
{\vec j}= {\vec J} - \sigma \vec V = \sum_j \sigma_j \vec V_j. \label{eq.8}
\end{equation}
To complete the set of Maxwell's equations, we quote that the magnetic field
must be divergence free
\begin{equation}
\frac{\partial}{\partial {\vec{\xi}}} \cdot \vec B^\prime = 0,
\label{eq.9}
\end{equation}
and that Faraday's induction equation requires the curl of the electric
field in the wave frame to vanish:
\begin{equation}
\frac{\partial}{\partial {\vec{\xi}}} \times \vec E^\prime = 0,
\label{eq.10}
\end{equation}
with the primed variables being defined in the wave frame. The Lorentz
transformation has been used to derive Eq.~(\ref{eq.10}), since it gives the
connection between the electromagnetic fields ($c$ is the speed of light in
vacuo) in the plasma's center of momentum frame and the moving wave frame
through the relation
\begin{equation}
\vec E^\prime = \vec E + \frac{1}{c} \vec V \times \vec B.
\label{eq.11}
\end{equation}
Of course, the magnetic field remains invariant to lowest order in $V/c$, and
thus $\vec B^\prime = \vec B$. For later purposes, we define the mass density
of particle kind $j$ as $\varrho_j = n_j m_j$, with the total mass density
\begin{equation} \varrho = \sum_j \varrho_j. \label{eq.12}
\end{equation}
The center of momentum velocity (for which we are free to choose $\vec{U}=0$)
is defined as follows:
\begin{equation}
\varrho \vec U = \sum_j \varrho_j \vec U_j.
\label{eq.13}
\end{equation}
Since we are interested in the individual ion and electron dynamics, we do
not sum their momentum equations like in MHD, but use
instead the separate multi-fluid equations. The individual momentum equation
of species $j$ can conveniently be quoted in conservation form in the moving
frame, reading
\begin{equation}
\frac{\partial}{\partial {\vec{\xi}}} \cdot \left( m_j n_j \vec V_j \vec V_j
+ p_j \mathbf{1} \right) = q_j n_j \left( \vec E^\prime + \frac{1}{c} \vec
V_j \times \vec B^\prime \right). \label{eq.14}
\end{equation}
The expression $\vec V_j\vec V_j$ means a tensor in dyadic notation, and
$\mathbf{1}$ the unit dyade. For the equation of the partial pressure, we may
take a simple polytropic equation of state for the purpose of closure, and
thus write
\begin{equation}
p_j = p_{j0} \left( \frac{n_j}{n_{j0}} \right)^{\gamma_j},
\label{eq.15}
\end{equation}
with some constant reference density, $n_{j0}$, and pressure, $p_{j0}$.
Equivalently, we may consider the (polytropic, with index $\gamma_j$) entropy
equation
\begin{equation}
\vec V_j \cdot \frac{\partial}{\partial {\vec{\xi}}} ln \left( p_j \varrho_j^{-\gamma_j}
\right) = 0.
\label{eq.16}
\end{equation}
The set of  Eqs.~(\ref{eq.2}), (\ref{eq.5}), (\ref{eq.7}), (\ref{eq.9}),
(\ref{eq.10}), (\ref{eq.11}), (\ref{eq.14}), and (\ref{eq.15}) is closed and sufficient
to calculate all the independent but coupled variables. In what follows we
shall assume a reduced geometry.

\subsection{Reduced multi-fluid equations in one-dimensional geometry}

We consider a one-dimensional spatial setup, i.e. a dependence on only one
spatial coordinate and take components with respect to $\hat{{\vec n}}$,
where the unit vector may correspond to the wave unit vector $\hat{{\vec k}}$
in Fourier variables. Thus we generally define the components
\begin{equation}
\xi = \hat{{\vec n}} \cdot ( {\vec x} - \vec V t ),
\label{eq.17}
\end{equation}
\begin{equation}
\vec V_j = V_{j\mathrm n} \hat{{\vec n}} + \vec V_{j \mathrm t},
\label{eq.18}
\end{equation}
\begin{equation}
\vec V_{j \mathrm t} = (\vec 1 - \hat{{\vec n}} \hat{{\vec n}} ) \cdot {\vec
V}_j.
\label{eq.19}
\end{equation}
Transverse components are obtained by projection perpendicular to the
longitudinal direction. The corresponding magnetic field components are
defined as $B_{\mathrm n} = \hat{{\vec n}} \cdot \vec B$ and $\vec B_{\mathrm
t} = \left( \underline{\underline{1}} - \hat{{\vec n}} \hat{{\vec n}} \right)
\cdot \vec B$, from which it follows that $\hat{{\vec n}} \times \vec
B_{\mathrm t} = \hat{{\vec n}} \times \vec B$, and that $\vec B_{\mathrm t}
\cdot \hat{{\vec n}}= 0$. The fluid equations then read as follows. For the
longitudinal momentum conservation, we have
\begin{equation}
\frac{\mathrm d}{\mathrm d \xi} \left( n_j V_{j\mathrm n} V_{j\mathrm n} + \frac{p_j}{m_j} \right) =
\frac{q_j n_j}{m_j} \left( E^\prime_{\mathrm n} + \frac{1}{c} \left( \vec V_j \times
\vec B^\prime \right) \cdot \hat{{\vec n}} \right), \label{eq.21}
\end{equation}
and for transverse momentum conservation we obtain the equation
\begin{equation}
\frac{\mathrm d}{\mathrm d \xi} \left( n_j V_{j\mathrm n} \vec V_{j \mathrm t} \right) = \frac{q_j n_j}{m_j}
\left( \vec E^\prime_{\mathrm t} + \frac{1}{c} \left( \vec V_j \times \vec B^\prime
\right)_{\mathrm t} \right). \label{eq.22}
\end{equation}
The longitudinal magnetic field is strictly conserved and thus constant:
\begin{equation}
\frac{\mathrm dB^\prime_{\mathrm n}}{\mathrm d \xi}  = 0, \label{eq.23}
\end{equation}
and the charge density obeys Gau\ss{}' law
\begin{equation}
\frac{\mathrm dE^\prime_{\mathrm n}}{\mathrm d \xi}  = 4\pi \sum_j q_j n_j. \label{eq.24}
\end{equation}
If we take after Eq.~(\ref{eq.10}) the curl of $\vec E^\prime$ in the wave frame, we find
\begin{equation}
\hat{{\vec n}} \frac{\mathrm d}{\mathrm d \xi} \times \vec E^\prime = \hat{{\vec n}} \times
\frac{\mathrm d \vec E^\prime_{\mathrm t} }{\mathrm d \xi} = 0 = \hat{{\vec n}} \times \frac{\mathrm d\vec E_{\mathrm t}}{\mathrm d \xi}
 + \hat{{\vec n}} \times \left( \frac{1}{c} V \hat{{\vec n}} \times
\frac{\mathrm d \vec B_{\mathrm t}}{\mathrm d \xi} \right), \label{eq.28}
\end{equation}
where the last part of the equation follows from the previous Eq.~(\ref{eq.11}). In conclusion, $\vec E^\prime_{\mathrm t}$ is constant and can
be set equal to zero. Using this result, the transverse electric field is
obtained through Eq.~(\ref{eq.11}), which is equivalent to writing
\begin{equation}
\vec E_{\mathrm t} = - \frac{1}{c} V ( \hat{{\vec n}} \times \vec B_{\mathrm t} ),
\label{eq.29}
\end{equation}
and makes the transverse electric field a dependent auxiliary variable being
fully determined by the transverse magnetic field. For the longitudinal electric
field component in the wave frame, one has
\begin{equation}
E^\prime_{\mathrm n} = E_{\mathrm n} + \frac{1}{c} \hat{{\vec n}} \cdot ( V \hat{{\vec n}} \times \vec B ),
\label{eq.30}
\end{equation}
which yields $E^\prime_{\mathrm n} = E_{\mathrm n}$, which is to be used in
Gau{\ss}' law (\ref{eq.24}). For the one-dimensional spatial geometry chosen
here, the longitudinal current density can be written as
\begin{equation}
J_{\mathrm n} = \frac{c}{4 \pi} \hat{{\vec n}} \cdot \left( \hat{{\vec n}} \times
\frac{\mathrm d\vec B}{\mathrm d \xi}  \right) + \frac{1}{4 \pi} \; \frac{\mathrm d}{\mathrm d \xi} (V E_{\mathrm n}).
\label{eq.25}
\end{equation}
Since the curl of $\vec B$ has only transverse components, we obtain from
Eq.~(\ref{eq.25}) that the longitudinal current density in the wave frame must be
strictly constant, since it is given by
\begin{equation}
j_{\mathrm n} = \sum_j q_j n_j V_{j\mathrm n} = \sum_j \sigma_j V_{j\mathrm n} = \sum_j q_j F_{j\mathrm n},
\label{eq.26}
\end{equation}
where the individual particles fluxes ($F_{j\mathrm n}$) are conserved according to the
longitudinal continuity equation, which expresses flux conservation in the
form
\begin{equation}
\frac{\mathrm d}{\mathrm d \xi} \left( n_j V_{j\mathrm n} \right) = \frac{\mathrm d F_{j\mathrm n}}{\mathrm d \xi} = 0.
\label{eq.38}
\end{equation}
The transverse component of Amp\`ere's law including the induction current
can be cast in the form
\begin{equation}
{\vec j}_{\mathrm t} = \frac{c}{4 \pi} \left( \hat{{\vec n}} \times \frac{\mathrm d{\vec
B}_{\mathrm t}}{\mathrm d \xi}  \right) + \frac{1}{4 \pi} V \frac{\mathrm d\vec E_{\mathrm t}}{\mathrm d \xi}  = \sum_j q_j n_j
\vec V_{j \mathrm t}, \label{eq.27}
\end{equation}
whereby we note that $\vec V_{j \mathrm t}=\vec U_{j \mathrm t}$, since $\vec
V$ has no transverse component, but $\vec V_{j\mathrm n}=\vec U_{j\mathrm
n}-V$. The right-hand side of Eq.~(\ref{eq.28}) gives an expression for the
gradient of the transverse electric field component. It can be inserted in
Amp\`ere's law, which thus can be written as
\begin{equation}
\frac{1}{4 \pi} V \frac{\mathrm d\vec E_{\mathrm t}}{\mathrm d \xi} = - \frac{1}{4 \pi c} V^2 \left(
\hat{{\vec n}} \times \frac{\mathrm d\vec B_{\mathrm t}}{\mathrm d \xi}  \right) = \sum_j n_jq_j {\vec
V}_{j \mathrm t} - \frac{c}{4 \pi} \left( \hat{{\vec n}} \times \frac{\mathrm d{\vec
B}_{\mathrm t}}{\mathrm d \xi}  \right). \label{eq.33}
\end{equation}
If $(V/c)^2 \ll 1$, which we will assume in the remainder, then the
displacement current term can be safely neglected. It is certainly needed if
one wants to make the transition to free electromagnetic waves correctly, in
which we are not interested here. We will therefore not keep this term
anymore. The basic equation for the magnetic field, which is Amp\`{e}re's law
for the transverse component, then reads
\begin{equation}
\frac{4 \pi}{c} \sum_j n_j q_j \vec V_{j \mathrm t} = \frac{\mathrm d}{\mathrm d \xi} \left(
\hat{{\vec n}} \times \vec B_{\mathrm t} \right). \label{eq.40}
\end{equation}
The magnetic field is free of divergence, which means in our geometry and
variables that Eq.~(\ref{eq.23}) is fulfilled with $B_{\mathrm n}=B_{\mathrm
n}^\prime$. The last two equations together fully determine the vector
magnetic field, given the current density is provided. The transverse
electric field is obtained from Eq.~(\ref{eq.29}) and the longitudinal one from
Gau{\ss}' law Eq.~(\ref{eq.24}) with $E_{\mathrm n}=E_{\mathrm n}^\prime$.

Next we quote again the transverse momentum equation for each species, which reads
\begin{equation}
\frac{\mathrm d}{\mathrm d \xi} \left( n_j V_{j\mathrm n} \vec V_{j \mathrm t} \right) = \frac{q_j}{m_j c}
n_j \left( \vec V_j \times \vec B \right)_{\mathrm t} = \frac{q_j n_j}{m_j c} \left(
V_{j\mathrm n} \hat{{\vec n}} \times \vec B_{\mathrm t} + \vec V_{j \mathrm t} \times \hat{{\vec n}} B_{\mathrm n}
\right), \label{eq.35}
\end{equation}
whereby the electric field term has been written out in detail. Similarly,
the longitudinal momentum equation reads
\begin{equation}
\frac{\mathrm d}{\mathrm d \xi} \left( n_j V^2_{j\mathrm n} + \frac{p_j}{m_j} \right) = \frac{q_j
n_j}{m_j} \left( E_{\mathrm n} + \frac{1}{c} (\vec V_j \times \vec B) \cdot \hat{{\vec
n}} \right), \label{eq.36}
\end{equation}
which must be supplemented to obtain closure by the entropy or pressure
equation,
\begin{equation}
\frac{\mathrm d}{\mathrm d \xi} ln \left( p_j \varrho_j^{-\gamma_j} \right) = 0. \label{eq.37}
\end{equation}

In what follows it turns out to be convenient to use the natural spatial and
temporal scales of the multicomponent plasma, which depend on the various
fluid and field parameters. We define a longitudinal gyration length for the
species $j$ as
\begin{equation}
r_j = \frac{V_{j\mathrm n}}{\Omega_j} = \frac{F_{j\mathrm n}}{\Omega_j} \; \frac{1}{n_j},
\label{eq.41}
\end{equation}
which corresponds to the gyration radius calculated with the longitudinal
velocity instead of the perpendicular one. It is implicitly dependent on
$n_j$ via the drift speed and Eq.~(\ref{eq.38}). Another interesting length is
given by the strictly constant quantity
\begin{equation}
L_j = \frac{B_{\mathrm n} c}{4 \pi q_j F_{j\mathrm n}} =
 r_j \left( \frac{V_{\mathrm Aj}}{V_{j\mathrm n}} \right)^2 =
\frac{1}{r_j} \left( \frac{c}{\omega_j} \right)^2.
\label{eq.42}
\end{equation}
Here we have introduced the Alfv\'{e}n speed, $V_{\mathrm Aj}^2 = B_{\mathrm
n}^2/(4 \pi n_j m_j)$, based on the mass density $\varrho_j$ of species $j$
only, and the respective plasma frequency $\omega_j^2 = 4 \pi q_j^2 n_j /
m_j$ and gyrofrequency $\Omega_j = B_{\mathrm n} q_j /(m_j c)$, carrying the
sign of the charge $q_j$. Note that $L_j$ is strictly constant, and its
inverse sums over all species up to zero, i.e. $\sum_j 1/L_j = 0$, because of
the condition of zero longitudinal total current: $j_{\mathrm n} - \sigma V =
\sum_j q_j n_j V_{j\mathrm n} = 0$. The standard Alfv\'{e}n speed based on
$B_{\mathrm n}$ is obtained by the summation
\begin{equation}
\sum_j \frac{1}{V^2_{\mathrm Aj}} = \frac{1}{V^2_{\mathrm A}}.
\label{eq.43}
\end{equation}
Concerning the compressive dynamics, it is important to note that
longitudinal and transverse motions are coupled through $p_j$ and $r_j =
V_{j\mathrm n} / \Omega_j$, i.e. through the particle number density, when
the mass continuity equation (\ref{eq.38}) and entropy equation (\ref{eq.37}) are
exploited.

\section{Wave equations}

In this section, we recast the basic equations for the fields and plasma
multi-fluid parameters into the form of coupled wave equations. For that
purpose, we shall not straightforwardly Fourier transform them but rather
rewrite them, by use of multiple differentiation, in such a form that we
finally obtain single ``wave equations'' for the electric field and magnetic field
components. Remember that in the moving frame all variables depend solely on
the coordinate $\xi = {\vec x} \cdot \hat{{\vec n}} - V t $. We first consider
the pressure equation again. We may use for the species' sound speed the
standard definition
\begin{equation}
c_{j}^2 = \frac{\partial p_j}{\partial \varrho_j },
\label{eq.44}
\end{equation}
and can then re-evaluate the momentum equation by use of
\begin{eqnarray}
\frac{\mathrm d}{\mathrm d \xi} \left( \frac{p_j}{m_j} \right) = c_{j}^2 \frac{\mathrm d n_j}{\mathrm d \xi},
\label{eq.45}
\end{eqnarray}
which together with Eq.~(\ref{eq.38}) allows us to quote the longitudinal
momentum equation in the form
\begin{equation}
\left( c^2_{j} - V^2_{j\mathrm n} \right) \frac{\mathrm d n_j}{\mathrm d \xi} = \frac{q_j n_j}{m_j}
\left( E_{\mathrm n} + \frac{1}{c} (\vec V_{j \mathrm t} \times \vec B_{\mathrm t}) \cdot
\hat{{\vec n}} \right). \label{eq.46}
\end{equation}
It is convenient to introduce the effective Debye length $\lambda_j$ of
species $j$ as follows:
\begin{equation}
\frac{1}{\lambda_{j}^2} = \frac{\omega^2_{j}}{c^2_{j}  - V^2_{j\mathrm n}},
\label{eq.47}
\end{equation}
the sum of which gives the total Debye length, still including the
differential drifts:
\begin{equation}
\frac{1}{\lambda_{\mathrm{D}}^2} = \sum_j \frac{1}{\lambda_{j}^2}.
\label{eq.48}
\end{equation}
Both $\lambda_j$ and $\lambda_{\mathrm D}$ are not necessarily real
quantities. Each species brings in its own length scale $\lambda_j$. It is
also convenient to introduce the second-order wave operator (which
parametrically depends still on $V$ via the $V_{j\mathrm n}$):
\begin{equation}
{\mathcal D}_E = \frac{\mathrm d^2}{\mathrm d \xi^2} - \frac{1}{\lambda_{\mathrm{D}}^2}.
\label{eq.49}
\end{equation}
Finally, one obtains a driven wave equation for the longitudinal electric field
\begin{equation}
{\mathcal D}_E \, E_{\mathrm n} = \frac{1}{c} \sum_j \frac{1}{\lambda^2_{j}}
(\vec V_{j \mathrm t} \times \vec B_{\mathrm t} ) \cdot \hat{{\vec n}},
\label{eq.50}
\end{equation}
in which the transverse particle motions and electromagnetic fields show up
through a nonlinear electromotive force, which is the summed contribution of
the longitudinal components of the Lorentz forces acting on each species.
This driving force acting on $E_{\mathrm n}$ resembles a ``convection''
electric field. When being decoupled from the transverse plasma and field
dynamics, the longitudinal electric field equation just describes free
electrostatic oscillations, such as Langmuir and acoustic waves, as we will
show later.

Let us return to the transverse momentum equation and rewrite it by exploiting
the mass continuity equation. Then it is straightforward to derive
\begin{equation}
\frac{\mathrm d\vec V_{j \mathrm t} }{\mathrm d \xi} = \frac{1}{r_j} \left( \frac{V_{j\mathrm n}}{B_{\mathrm n}} \;
\hat{{\vec n}} \times \vec B_{\mathrm t} - \hat{{\vec n}} \times \vec V_{j \mathrm t} \right).
\label{eq.51}
\end{equation}
Using this equation, we can rewrite the normal component of the convection
electric field which occurs in Eqs.~(\ref{eq.46}) and (\ref{eq.50}) as
follows:
\begin{equation}
E_{j\mathrm n} \equiv \frac{1}{c}\left(\vec V_{j \mathrm t} \times \vec B_{\mathrm t}\right) \cdot \hat{{\vec
n}} = - \frac{B_{\mathrm n}}{V_{j\mathrm n}} \frac{r_j}{c} \frac{\mathrm d}{\mathrm d \xi}
\left(\frac{1}{2}\vec{V}_{j \mathrm t}^2\right) = - \frac{m_j}{q_j} \frac{\mathrm d}{\mathrm d \xi}
\left(\frac{1}{2}\vec{V}_{j \mathrm t}^2\right), \label{eq.51a}
\end{equation}
a relation which is going to be used later. If the module of the transverse
plasma velocity of species $j$ is constant, then its convection electric
field $E_{j\mathrm n}$ vanishes. Using Amp\`{e}re's law Eq.~(\ref{eq.40}) and the
previous Eq.~(\ref{eq.51a}), we can derive by vector cross multiplication
of Eq.~(\ref{eq.40}) with $\hat{{\vec n}}$ and subsequent scalar multiplication
with $\vec{B}_{\mathrm t}$ another conservation law:
\begin{equation}
\frac{\mathrm d}{\mathrm d \xi} \left( \sum_j \varrho_j \vec{V}_{j \mathrm t}^2 - \frac{1}{4 \pi}
\vec{B}_{\mathrm t}^2 \right) = 0. \label{eq.51b}
\end{equation}
If the integration constant is zero, this equation expresses equipartition
between the transverse total particle kinetic energy and the transverse
magnetic energy, like it is the case in a classical MHD Alfv\'en wave.

We consider now the equation for the transverse magnetic field. By
differentiation of Amp\`{e}re's law, we can obtain a second-order nonlinear
wave equation for the transverse magnetic field
\begin{equation}
\frac{\mathrm d^2}{\mathrm d \xi^2} \, \hat{{\vec n}} \times \vec B_{\mathrm t} = \frac{4 \pi}{c}
\sum_j q_j \left( n_j \frac{\mathrm d\vec V_{j \mathrm t} }{\mathrm d \xi}+ \vec V_{j \mathrm t} \frac{\mathrm dn_j }{\mathrm d
\xi} \right). \label{eq.52}
\end{equation}
Here, it is convenient to introduce the skin depth or inertial length $\ell_j$
of species $j$ defined as follows:
\begin{equation}
\frac{1}{\ell_j^2} = \frac{\omega^2_{j}}{c^2},
\label{eq.53}
\end{equation}
the sum of which gives the total skin depth:
\begin{equation}
\frac{1}{\ell_{\mathrm{S}}^2} = \sum_j \frac{1}{\ell_j^2},
\label{eq.54}
\end{equation}
where each species brings in its own length scale $\ell_j$. It is again
convenient to introduce a second-order wave operator
\begin{equation}
{\mathcal D}_B = \frac{\mathrm d^2}{\mathrm d \xi^2} - \frac{1}{\ell_{\mathrm{S}}^2}. \label{eq.55}
\end{equation}
Using this, one finally obtains a driven wave equation for the transverse
magnetic field
\begin{equation}
{\mathcal D}_B \, \vec{B}_{\mathrm t} = -\sum_j \left( \frac{B_{\mathrm n}}{V_{j\mathrm n}}\frac{1}{\ell_j^2}\vec V_{j \mathrm t}
+ \frac{1}{\lambda_{j}^2 \,c} ( \hat{{\vec n}} \times \vec V_{j \mathrm t})
\left[ E_{\mathrm n} + \frac{1}{c} (\vec V_{j \mathrm t} \times \vec B_{\mathrm t}) \cdot \hat{{\vec n}}
\right] \right).
\label{eq.56}
\end{equation}
On the right-hand side, the transverse currents appear and the longitudinal
charge density variations and related electrostatic effects show up through
the nonlinear electromotive force, which involves the longitudinal electric
field. Note that this nonlinear driver contains the natural length scales
(density dependent) of all the species involved. Of course, this wave
equation seems, without further approximation, quite formal but elucidates
the nature of the coupling of the unforced transverse magnetic field
(dynamics described by the operator ${\mathcal D}_B $) with the compressive
electrostatic fluctuations and transverse plasma motions. When being
decoupled from the plasma currents and electric-field (no charges) dynamics,
this transverse magnetic field equation just describes the finite penetration
of the magnetic field into the skin layer of the plasma and results in its
exponential decline on the length scale $\ell_{\mathrm{S}}$. To gain better
insight into the terms contributing to Eq.~(\ref{eq.56}), we may rewrite
it also in the form
\begin{equation}
{\mathcal D}_B \, \vec{B}_{\mathrm t} =- \sum_j  \frac{\omega^2_{j}}{c^2} \left( \frac{B_{\mathrm n}}{V_{j\mathrm n}}\vec V_{j \mathrm t}
+ \frac{ c(E_{\mathrm n} + E_{j\mathrm n}) }{c^2_j  - V^2_{j\mathrm n}} (\hat{{\vec n}} \times \vec V_{j \mathrm t}) \right).
\label{eq.57a}
\end{equation}
Similarly, one can also rewrite the driven wave equation for the longitudinal
electric field in the concise form
\begin{equation}
{\mathcal D}_E \, E_{\mathrm n} = \sum_j \frac{1}{\lambda^2_{j}} E_{j\mathrm n},
\label{eq.57b}
\end{equation}
where we remind that $E_{j\mathrm n}$ can be derived after Eq.~(\ref{eq.51a})
from a potential that is given by the transverse kinetic energy of species
$j$. These two coupled nonlinear equations are completed and closed by Eq.~(\ref{eq.46}) for the density and Eq.~(\ref{eq.51}) for the transverse
velocity of each species.

So far, we made neither any approximation nor linearization, but just
inserted the original momentum equations (\ref{eq.46}) and (\ref{eq.51}) into
the differentiated Amp\`{e}re's and Gau{\ss}' laws. Therefore, the above
equations depend in a highly nonlinear manner still on the different number
densities $n_j$. Yet the transition in Eqs.~(\ref{eq.57a}) and (\ref{eq.51}) to
the incompressible limit is simple, because then the electrostatic nonlinear
forcing terms $E_{\mathrm n}$ and $E_{j\mathrm n}$ vanish, and the plasma
frequency $\omega_j$ and the inertial length $\ell_j$, respectively, become
constants as defined by the fixed background density of species $j$. We
discuss this in the section after the next one.

\section{Eigenmodes and driven waves}

\subsection{Nonlinear Alfv\'en-ion-cyclotron waves}

In order to maintain the linear form of the original equations, it is
convenient to introduce new variables, relating to left- and right-hand
polarized fields, which are defined as follows:
\begin{equation}
\vec B^\pm_{\mathrm t} = \vec B_{\mathrm t} \pm \left( \hat{{\vec n}} \times \vec B_{\mathrm t} \right),
\label{eq.58}
\end{equation}
\begin{equation}
\vec V^\pm_{\mathrm t} = \vec V_{\mathrm t} \pm \left( \hat{{\vec n}} \times \vec V_{\mathrm t} \right).
\label{eq.59}
\end{equation}
These field variables are orthogonal, i.e. $\vec B^\pm_{\mathrm t} \cdot
\vec B^\mp_{\mathrm t} =0$, and $\vec V^\pm_{j \mathrm t} \cdot \vec V^\mp_{j
\mathrm t} =0$. By taking the cross product of Eqs.~(\ref{eq.40}) and
(\ref{eq.51}) with the unit vector $\hat{{\vec n}}$, we obtain after some
algebra the equations of motion for the circular transverse variables:
\begin{equation}
\frac{\mathrm d\vec B^\pm_{\mathrm t}}{\mathrm d \xi}  = \pm \frac{4 \pi}{c} \sum _j \sigma_j {\vec
V}^\mp_{j \mathrm t}, \label{eq.60}
\end{equation}
\begin{equation}
\frac{\mathrm d\vec V^\pm_{j \mathrm t}}{\mathrm d \xi}  = \mp \frac{1}{r_j} \left(
\frac{V_{j\mathrm n}}{B_{\mathrm n}} \vec B^\mp_{\mathrm t} - \vec V^\mp_{j \mathrm t} \right). \label{eq.61}
\end{equation}

Let us now first consider nonlinear incompressible solutions. Since
$E_{\mathrm n}=0$, consequently quasineutrality strictly holds, $\sum_j q_j
n_j=0$. The velocity fields and magnetic field must have constant modules and
be aligned, so that after Eq.~(\ref{eq.51a}) their respective vector cross
product, and thus also $E_{j\mathrm n}$ vanishes. As $n_j$, $r_j$,
$V_{j\mathrm n}$, $c_j$, and $\omega_j$ then all are constant, we can solve
the resulting linear set Eqs.~(\ref{eq.60}) and (\ref{eq.61}) by Fourier
tansformation (FT), without putting any limitations on the amplitudes of
$\vec B^\pm_{\mathrm t}$ or $\vec V^\pm_{j \mathrm t}$, other than from
Eq.~(\ref{eq.51b}) which implies that the magnetic field amplitude is also
constant. As usually, FT means that $\mathrm d/\mathrm d \xi
\rightarrow i\,k$, and thus we can invert the transverse momentum equation
(\ref{eq.51}), which yields with the normalized wave vector,
$\kappa_j=kV_{j\mathrm n}/\Omega_j$, the complex vector relation
\begin{equation}
\tilde{\vec{V}}_{j \mathrm t}(k) = \frac{V_{j\mathrm n}}{B_{\mathrm n}(1-\kappa_j^2)}\left( \tilde{\vec{B}}_{\mathrm t}(k)
+ i \kappa_j \, \hat{{\vec n}} \times \tilde{\vec{B}}_{\mathrm t}(k) \right).
\label{eq.64}
\end{equation}
This result can be inserted into the FT of Eq.~(\ref{eq.57a}) to obtain the
algebraic wave equation
\begin{equation}
\left(k^2 + \frac{1}{\ell_{\mathrm{S}}^2}\right) \tilde{\vec{B}}_{\mathrm t}(k) = \sum_j \frac{1}{\ell_j^2} \frac{1}{1-\kappa_j^2} \left( \tilde{\vec{B}}_{\mathrm t}(k)
+ i \kappa_j ( \hat{{\vec n}} \times \tilde{\vec{B}}_{\mathrm t}(k) ) \right),
\label{eq.65}
\end{equation}
which may also be written as
\begin{equation}
\left( k^2 - \sum_j \frac{1}{\ell_j^2} \frac{\kappa_j^2}{1-\kappa_j^2} \right)
\tilde{\vec{B}}_{\mathrm t}(k) = i \left(\sum_j \frac{1}{\ell_j^2} \frac{\kappa_j
}{1-\kappa_j^2} \right)( \hat{{\vec n}} \times \tilde{\vec{B}}_{\mathrm t}(k) ),
\label{eq.66}
\end{equation}
and which yields, by taking the vector cross product of Eq.~(\ref{eq.66}) with
$\hat{{\vec n}}$ and by resolving the resulting two equations, the two
dispersion relations describing left- and right-hand polarized waves as
follows:
\begin{equation}
k^2 = \sum_j \left(\frac{\omega_j}{c}\right)^2\frac{ \pm \kappa_j }{1\mp \kappa_j} =
\sum_j \hat{\varrho}_j \left(\frac{\Omega_j}{V_{\mathrm A}}\right)^2 \frac{ \pm \kappa_j }{1\mp
\kappa_j},
\label{eq.67}
\end{equation}
with the fractional mass density $\hat{\varrho}_j= \varrho_j/\varrho$. Eq.~(\ref{eq.67}) is nothing else but the standard dispersion relation
for the Alfv\'en/ion-cyclotron (and magnetosonic-whistler) waves in a
multicomponent plasma with the differential drifts contained in $V_{j\mathrm
n}$ and for parallel propagation \citep[see e.g.][]{davidson83}, yet which
applies here to arbitrarily large wave amplitudes. Apparently, the wave
frequency is obtained by the Doppler shift formula, $\omega = kV$, where $V$
is hidden in $V_{j\mathrm n}= U_{j\mathrm n}-V$, i.e. in
$\kappa_j=kV_{j\mathrm n}/\Omega_j$. Once the wave vector $k=k(V)$ is known,
the wave frequency is obtained as a function of the phase speed $V$.

Let us consider the long-wavelength limit of Eq.~(\ref{eq.67}), which when being
expanded to second order in $\kappa_j$ reads
\begin{equation}
1 = \sum_j \hat{\varrho}_j \left(\frac{\Omega_j}{kV_{\mathrm A}}\right)^2 (\pm \kappa_j)(1 \pm \kappa_j),
\label{eq.68}
\end{equation}
where the first term of the sum vanishes, since
$\sum_j\left(\frac{\omega_j^2}{\Omega_j}\right)V_{j\mathrm n}=0$ because of
the quasi-neutrality condition, i.e. $\sigma = 0$ in Eq.~(\ref{eq.4}), and the
zero-longitudinal-current constraint (\ref{eq.26}). The second term then
yields
\begin{equation}
1 = \sum_j \hat{\varrho}_j \frac{(U_{j\mathrm n}^2-2U_{j\mathrm n}V + V^2)}{V_{\mathrm A}^2}.
\label{eq.69}
\end{equation}
As a vanishing bulk speed, $\vec{U}=0$, may be assumed, the
center-of-momentum condition means that $\sum_j \hat{\varrho}_j U_{j\mathrm
n} = 0$, and thus we can solve for the phase speed in the center of momentum
frame and obtain
\begin{equation}
V = \pm V_{\mathrm A} \sqrt{1 - \sum_j \hat{\varrho}_j \left(\frac{U_{j\mathrm n}}{V_{\mathrm A}}\right)^2}\,.
\label{eq.70}
\end{equation}
This is the phase speed of an Alfv\'en wave in a multicomponent plasma,
including field-aligned drift motions, leading to a slowing down of the phase
speed.

We shall now derive the incompressible Alfv\'en-cyclotron wave without resort
to the FT, but recourse instead on Eq.~(\ref{eq.51a}). Since
$E_{j\mathrm n}=0$, each velocity vector and magnetic field must be aligned,
which generally implies that $\vec{V}_{j \mathrm t} = a_j \vec{B}_{\mathrm
t}$. This can be inserted in Eq.~(\ref{eq.51}) to obtain
\begin{equation}
\frac{\mathrm d\vec{V}_{j \mathrm t}}{\mathrm d \xi}  = \frac{1}{r_j} \left(\frac{V_{j\mathrm n}}{B_{\mathrm n} a_j} -1\right)\,
\hat{{\vec n}} \times \vec V_{j \mathrm t}. \label{eq.75}
\end{equation}
Twofold differentiation yields the simple harmonic oscillator equation for
the gyromotion:
\begin{equation}
\left(\frac{\mathrm d^2}{\mathrm d \xi^2} + k_j^2\right) \vec{V}_{j \mathrm t} = 0, \label{eq.76a}
\end{equation}
with the squared wave vector defined as
\begin{equation}
k_j^2 = \frac{1}{r_j^2}\left(\frac{V_{j\mathrm n}}{B_{\mathrm n} a_j} -1\right)^2.
\label{eq.76b}
\end{equation}
Since all species must spatially oscillate in the same way, the wave vector
must not depend on the index $j$, i.e. we can put $k_j= \pm k$, which yields
with $\kappa_j = k r_j$ two possible solutions for the desired
proportionality coefficient:
\begin{equation}
a_j = \frac{V_{j\mathrm n}}{B_{\mathrm n}} \frac{1}{1 \pm \kappa_j}.
\label{eq.76c}
\end{equation}
Knowing the coefficient $a_j$, we can use it in the wave equation (\ref{eq.57a})
without electric fields and get another harmonic oscillator equation:
\begin{equation}
\left(\frac{\mathrm d^2}{\mathrm d \xi^2} + q^2\right) \vec{B}_{\mathrm t} = 0, \label{eq.77a}
\end{equation}
where the squared wave vector $q$ is an abbreviation for exactly the same sum
as appearing on the right hand side of Eq.~(\ref{eq.67}). Since all velocities
and the magnetic field are aligned, the wave vector $q$ must be equal to $k$,
and thus we again obtain the same dispersion relation as in the previous
Fourier analysis. Finally, the polarization relation (with the plus sign for
incompressible Alfv\'en-cyclotron and minus for magnetosonic-whistler waves)
reads
\begin{equation}
\vec{V}_{j \mathrm t} = \frac{V_{j\mathrm n}}{B_{\mathrm n}} \frac{1}{1 \pm \kappa_j} \vec{B}_{\mathrm t} .
\label{eq.77b}
\end{equation}
With this result, we can further evaluate the conservation law (\ref{eq.51b}),
and after some algebra obtain the result that
\begin{equation}
\frac{\mathrm d}{\mathrm d \xi} \left( \sum_j \hat{\varrho}_j V_{j\mathrm n}^2 \frac{1}{(1 \pm
\kappa_j)^2} - V_{\mathrm A}^2 \right) = 0. \label{eq.78a}
\end{equation}
Note that the dispersion relation (\ref{eq.67}) can also be cast into the form
\begin{equation}
V_{\mathrm A}^2 =  \sum_j \hat{\varrho}_j V_{j\mathrm n}^2 \frac{\mp\kappa_j}{\kappa_j^2(1 \pm
\kappa_j)},
\label{eq.78b}
\end{equation}
which facilitates a comparison with the previous equation. Expansion of
Eqs.~(\ref{eq.78a}) and (\ref{eq.78b}) to lowest order in $\kappa_j=kr_j$ yields
the MHD dispersion relation (\ref{eq.69}), i.e. in this case, the
conservation equation (\ref{eq.78a}) has a zero integration constant and
simply expresses equipartition between kinetic and magnetic energy densities.
This is not true any more if finite gyrokinetic effects are considered.

\subsection{Linear electrostatic waves}

Let us now discuss pure linear electrostatic waves, which are obtained by
taking the trivial solutions, $ \vec{V}_{j \mathrm t}= \vec{B}_{\mathrm t}=
0$, of the wave equation (\ref{eq.57a}), which also implies that
$E_{j\mathrm n}=0$. Then the linearized electrostatic wave equation
(\ref{eq.57b}) simply reads ${\mathcal D}_E \, E_{\mathrm n} =0$. After FT, we
obtain that $(k^2 + \lambda_{\mathrm{D}}^{-2}) \tilde{E}_{\mathrm n}(k) = 0$,
which explicitly yields the dispersion relation
\begin{equation}
k^2 = \sum_j \frac{\omega_j^2}{(U_{j\mathrm n}-V)^2- c_j^2}.
\label{eq.71}
\end{equation}
We may only consider here the case of zero drifts, i.e. $U_{j\mathrm n}=0$, and
a simple electron-proton plasma. Then we always find two solutions for $V^2$
from the equation
\begin{equation}
k^2 = \frac{\omega_{\mathrm e}^2}{V^2- c_{\mathrm e}^2} +
\frac{\omega_{\mathrm p}^2}{V^2- c_{\mathrm p}^2}. \label{eq.72}
\end{equation}
In the long-wavelength limit $(k \rightarrow 0)$, the diverging phase speed
$V_{\mathrm{L}}(k) = \omega_{\mathrm{P}}/k$ corresponds to the Langmuir wave,
with the total plasma frequency being defined by $\omega_{\mathrm{P}}^2=
\omega_{\mathrm e}^2+ \omega_{\mathrm p}^2$. For the ion-acoustic or sound
wave, we obtain the constant speed defined as
\begin{equation}
V_{\mathrm{S}} = \sqrt{ \frac{\omega_{\mathrm e}^2c_{\mathrm p}^2 + \omega_{\mathrm p}^2c_{\mathrm e}^2}{\omega_{\mathrm{P}}^2}} =
\sqrt{ \frac{k_{\mathrm{B}}(\gamma_{\mathrm e}T_{\mathrm e} + \gamma_{\mathrm p}T_{\mathrm p})}{m_{\mathrm e} + m_{\mathrm p}} }\,.
\label{eq.73}
\end{equation}
Here $k_{\mathrm{B}}$ is Boltzmann's constant. The general solution in terms of
frequency follows from the biquadratic equation
\begin{equation}
\omega^4 - \omega^2\left(\omega_{\mathrm e}^2+\omega_{\mathrm p}^2 + (c_{\mathrm e}k)^2 + (c_{\mathrm p}k)^2\right)
+ (c_{\mathrm e}k)^2(c_{\mathrm p}k)^2 + \omega_{\mathrm e}^2(c_{\mathrm p}k)^2 + \omega_{\mathrm p}^2(c_{\mathrm e}k)^2 = 0.
\label{eq.74}
\end{equation}
In the short-wavelength limit $(k \rightarrow \infty)$, we obtain two
solutions corresponding to the proton-acoustic wave with $\omega \approx k
c_{\mathrm p}$, or electron-acoustic wave with $\omega \approx k c_{\mathrm
e}$. Both modes are usually strongly Landau damped if a thermal Vlasov
description of the plasma is used. The sound wave, however, can exist since
$m_{\mathrm e} \ll m_{\mathrm p}$, and thus $V_{\mathrm{S}} \approx \sqrt{
k_{\mathrm{B}}\gamma_{\mathrm e}T_{\mathrm e}/ m_{\mathrm p}}$ for
$T_{\mathrm e} > T_{\mathrm p}$, so that strong proton Landau damping can be
avoided.

For a multicomponent plasma with drifts, the structure of the eigenmodes
becomes correspondingly richer, as each species contributes its own plasma
frequency and thermal speed, as well as drift speed. In the presence of a
compressive transverse wave, these modes all become coupled and are driven by
the nonlinear ponderomotive electric fields $E_{j\mathrm n}$ according to
Eq.~(\ref{eq.57b}). Similarly, the transverse eigenmodes defined by Eq.~(\ref{eq.67})
are driven according to Eq.~(\ref{eq.57a}) by the longitudinal electric field
$E_{\mathrm n}$ and the combined action of the various $E_{j\mathrm n}$.

\subsection{Compressive Alfv\'en-cyclotron--acoustic waves}

In this section, we will consider the coupling between the nonlinear
electromagnetic Alfv\'en-cyclotron waves and the electrostatic modes. We
recall that no assumptions, such as incompressibility, had to be made as to
derive the wave equations (\ref{eq.57a}) and (\ref{eq.57b}). We shall also
rewrite the transverse momentum equation (\ref{eq.51}), which describes the
gyromotion as a second-order wave equation. Since the longitudinal gyration
scale $r_j$ depends on the density according to Eq.~(\ref{eq.41}), its
differentiation has to be considered. If we neglect the magnetic field for a
moment, then we get for the transverse motion an equation in the form
\begin{equation}
\left( \frac{\mathrm d^2}{\mathrm d \xi^2} + \frac{1}{r_j^2} - \frac{\mathrm d \ln{n_j}}{\mathrm d \xi}
\frac{\mathrm d}{\mathrm d \xi} \right)\vec V_{j \mathrm t}=0. \label{eq.95}
\end{equation}
Mathematically speaking, this is the well known equation for a harmonic
oscillator, with an amplitude that may vary exponentially in $\xi$ at a scale
set by the density gradient length. For the differential operator yielding
harmonic oscillations (first two terms of Eq.~(\ref{eq.95})), we introduce the
symbol ${\mathcal D}_{V_j}$ to be used below. For the density gradient term,
we may approximately write
\begin{equation}
\frac{\mathrm d}{\mathrm d \xi}\ln{n_j}= \frac{\mathrm d}{\mathrm d \xi}
\ln{(\bar{n}_j + \delta n_j)} \approx \frac{1}{\bar{n}_j}\frac{\mathrm d}{\mathrm d \xi}
\delta n_j = \frac{\delta n_j}{\bar{n}_j} \frac{\mathrm d\ln{\delta n_j}}{\mathrm d \xi}
\end{equation}
with the average constant background density $\bar{n}_j$. If the gradient is
positive (negative), and therefore the density increases (decreases), the
longitudinal scale and thus the amplitude of $\vec V_{j \mathrm t}$ will
decrease (increase) correspondingly. However, as long as the relative density
variation remains small, a few percent say, this change will occur on a much
larger scale than $r_j$, namely $\bar{L}_j=r_j\bar{n}_j/\delta n_j$. If the
density fluctuates about zero, the net effect of the density modulation on
$\vec V_{j \mathrm t}$ will remain comparatively small.

For the sake of consistency, we will, in the remainder of this section, fully
retain this density-induced possible amplitude variation of $\vec V_{j
\mathrm t}$ according to Eq.~(\ref{eq.95}), but later on neglect the density
variations and consider all density-dependent parameters to be fixed at their
background values without denoting them explicitly by a barred symbol. Yet,
remember that the essential and lowest-order variations of the densities of
all species have been considered and already taken care of in Gau{\ss}' law
and the dynamics of $E_{\mathrm n}$, which indeed is of order unity as the
background electric field is zero, and similarly in Amp\`{e}re's law through
the appearance of the electric fields $E_{\mathrm n}$ and $E_{j\mathrm n}$.

We introduce, by using conserved or constant quantities, normalized
variables, such that $\vec{v}_{j \mathrm t}=\vec{V}_{j \mathrm t}/V$ and
$\vec{b}_{\mathrm t}=\vec{B}_{\mathrm t}/B_{\mathrm n}$. Similarly, we
introduce normalized electric fields as follows:
\begin{equation}
e_{j\mathrm n}=\frac{cE_{j\mathrm n}}{B_{\mathrm n}V},  \; \; \; e_{\mathrm n}=\frac{cE_{\mathrm n}}{B_{\mathrm n}V}.
\label{eq.77}
\end{equation}
We recall Eq.~(\ref{eq.57a}), according to which $e_{j\mathrm n}$ is
just an abbreviation for the gradient of a potential given by the transverse
kinetic energy, for which we have in normalized form
\begin{equation}
e_{j\mathrm n}= - \frac{V}{\Omega_j} \frac{\mathrm d}{\mathrm d \xi} \left(\frac{1}{2}\vec{v}_{j \mathrm t}^2\right).
\label{eq.88a}
\end{equation}
Using the same normalization for the transverse electric field (\ref{eq.29}),
we obtain simply that
\begin{equation}
\vec{e}_{\mathrm t}= - (\hat{{\vec n}} \times \vec{b}_{\mathrm t}),
\label{eq.88b}
\end{equation}
which is fully determined by the solution for $\vec{b}_{\mathrm t}$. Like
$\vec{e}_{\mathrm t}$, the charge densities $\sigma_j$ are now merely
auxiliary quantities and obtained from an integration of the previous
equation (\ref{eq.46}), which in the new variables can be written as
\begin{equation}
r_j \frac{\mathrm d}{\mathrm d \xi} \ln{\sigma_j} = (e_{\mathrm n} + e_{j\mathrm n})\frac{V V_{j\mathrm n}}{c^2_j -
V^2_{j\mathrm n}}, \label{eq.80}
\end{equation}
and then be formally integrated with the result
\begin{equation}
\sigma_j(\xi)= q_j\bar{n}_j \exp \left( V \Omega_j \int\limits_{\bar{\xi}}^{\xi}
\mathrm d\xi^\prime \,\frac{e_{\mathrm n}(\xi^\prime) + e_{j\mathrm n}(\xi^\prime)}{c^2_j(\xi^\prime) -
V^2_{j\mathrm n}(\xi^\prime)} \right). \label{eq.81}
\end{equation}

As an outcome of all the above considerations, we can now summarize the
resulting set of fluid wave equations. Firstly, we obtain for each species'
transverse motion a forced and amplitude-modulated harmonic oscillator
equation reading
\begin{equation}
{\mathcal D}_{V_j}\, \vec{v}_{j \mathrm t} = \left( \frac{V^2}{c^2_j - V^2_{j\mathrm n}}(e_{\mathrm n} +
e_{j\mathrm n}) \frac{\mathrm d{\vec v}_{j \mathrm t}}{\mathrm d \xi} + \frac{1}{r_j} \vec{b}_{\mathrm t} + \frac{\mathrm d}{\mathrm d
\xi}\hat{{\vec n}} \times \vec{b}_{\mathrm t} \right) \frac{\Omega_j}{V}.
\label{eq.76}
\end{equation}
Secondly, one can rewrite the mutually driven and coupled wave equations for
the longitudinal electric field and transverse magnetic field in the concise
forms
\begin{equation}
{\mathcal D}_E \, e_{\mathrm n} = \sum_j \frac{1}{\lambda^2_{j}} e_{j\mathrm n},
\label{eq.79a}
\end{equation}
\begin{equation}
{\mathcal D}_B \, \vec{b}_{\mathrm t} =- \sum_j \frac{1}{\ell_j^2} \left(  \frac{V}{V_{j\mathrm n}}
\vec{v}_{j \mathrm t} + \frac{V^2}{c^2_j - V^2_{j\mathrm n}}(e_{\mathrm n} + e_{j\mathrm n}) (\hat{{\vec n}}
\times {\vec v}_{j \mathrm t}) \right). \label{eq.79b}
\end{equation}

We recall again that up to this point of our algebraic derivations, no
linearization has been made, and the density variations were entirely
accounted for. Eqs.~(\ref{eq.80}) or (\ref{eq.81}) permit to calculate
the density of species $j$ completely through the line integral over the
electric fields that appear in the exponential Boltzmann factor in
Eq.~(\ref{eq.81}). So the density is a functional of the electric potentials.
However, this dependence of $n_j$ on $\xi$ may now without loss of essential
physics be neglected, if the density fluctuations can be assumed to remain
small (i.e. we do not want to consider MHD shocks \citep{goossens03}, or
electrostatic shocks and double layers here). Thus, all scales and parameters
such as $V_{j\mathrm n}, c_j, \omega_j, \lambda_j, \ell_j$, and $r_j$, which
have non-vanishing mean values, will, from here on, be calculated by use of the
background number density $\bar{n}_j$, as well as the conditions for
quasi-neutrality and zero longitudinal current and the center-of-momentum
condition. All compressive effects are described by the longitudinal electric
field $e_{\mathrm n}$ in this approximation. Consequently, the nonlinear
equations (\ref{eq.76}), (\ref{eq.79a}), and (\ref{eq.79b}) form a closed set,
which yet will generally require a numerical treatment to obtain solutions.
This is a task beyond the scope of this analytical paper.

Note that in each of these equations, the spatial variations are determined by
the natural scales of the dynamics of the involved field variables, i.e. by
the longitudinal scale $r_j$ for the transverse motions of the particles,
their Debye lengths $\lambda_j$ for the charge fluctuations, respectively,
their skin depths $\ell_j$ for the magnetic field penetration into the plasma
driven by the transverse currents. Differential motion of the species $j$
might be important and is therefore included in its drift speed $V _{j\mathrm
n}$. Its effect on the parametric instabilities has been studied in work
addressing the modulational and decay instability of Alfv{\'e}n waves by
considering streaming of alpha particles in the solar wind
\citep{hollweg93}. If there are no differential motions along the mean
field in the background plasma, i.e. if for all $j$ we have $U_{j\mathrm
n}=0$, then $V_{j\mathrm n}=-V$, and thus the longitudinal gyration scale
simply becomes $r_j=-V/\Omega_j$, which by its definition is not for each
species a positive definite quantity as the gyrofrequency carries the sign of
the charge of the species considered. The factor in front of the electric
field term in the above Eqs.~(\ref{eq.76}) and (\ref{eq.79b}) thus
changes in the drift-free case to $1/(1-(c_j/V)^2)$, which becomes
$1/(1-\beta_j)$ for $V=V_{\mathrm A}$, with the species plasma beta being
defined as $\beta_j=(c_j/V_{\mathrm A})^2$. For the parallel propagation
considered here, only this factor contains the thermal speed, and therefore
this factor simply becomes unity for a cold multi-species plasma without
drifts.

\subsection{Electric field fluctuations driven by an elliptically polarized Alfv\'en wave} \label{sect_ellipt}

In this section, we consider the electrostatic waves, which can be generated by
an elliptically polarized Alfv\'en wave, which for the sake of simplicity
we assume to be given. We shall then study the effect this wave has in
generating compressive fluctuations driven by the spatial variation of the
kinetic energy of the particles moving coherently in the same wave magnetic
field. The starting point is Eq.~(\ref{eq.79a}), which can with
the help of Eq.~(\ref{eq.88a}) be written as a driven oscillator equation for
the longitudinal electric field:
\begin{equation}
\mathcal D_E\, e_{\mathrm n}
= - \sum_j \frac{V}{2 \Omega_j} \frac{1}{\lambda_j^2} \frac{\mathrm d\vec{v}_{j \mathrm t}^2}{\mathrm d\xi}.
\label{eq.90}
\end{equation}
Before we write down the wave fields, let us define a proper coordinate
system. For the right-handed orthogonal system, we choose the unit vectors
$\hat{\vec{n}}=\vec{e}_3=\vec{e}_1\times\vec{e}_2$,
$\vec{e}_1=\vec{e}_2\times\hat{\vec{n}}$, and
$\vec{e}_2=\hat{\vec{n}}\times\vec{e}_1$. The wave may have a
wave vector $k$, and its normalized (dimensionless) magnetic field reads
\begin{equation}
\vec{b}_{\mathrm t}= b_1\vec{e}_1\cos{(k\xi)} +  b_2\vec{e}_2\sin{(k\xi)}.
\label{eq.91}
\end{equation}
Similarly, the related flow velocity of species $j$ is given by
\begin{equation}
\vec{v}_{j \mathrm t}= v_{j1}\vec{e}_1\cos{(k\xi)} +  v_{j2}\vec{e}_2\sin{(k\xi)}.
\label{eq.92}
\end{equation}
The associated electric field according to Eq.~(\ref{eq.90}) reads
\begin{equation}
e_{j\mathrm n}= \frac{kV}{\Omega_j} \cos{(k\xi)}\sin{(k\xi)}\left(v_{j1}^2 - v_{j2}^2\right).
\label{eq.93}
\end{equation}
On the other hand, using the original definition (\ref{eq.51a}) of this field,
we obtain the result
\begin{equation}
e_{j\mathrm n}= \cos{(k\xi)}\sin{(k\xi)}(b_2v_{j1} - b_1v_{j2}),
\label{eq.94}
\end{equation}
which by comparison of the last two equations, determines the particle velocity
components as $v_{j1,2} = b_{2,1}\,\Omega_j/(kV)$. Finally, we have
\begin{equation}
e_{j\mathrm n}= \frac{\Omega_j}{2kV}\sin{(2k\xi)}\left(b_1^2 - b_2^2\right),
\label{eq.96}
\end{equation}
which can be inserted in Eq.~(\ref{eq.79a}). We obtain the forced
oscillator equation
\begin{equation}
\frac{\mathrm d^2e_{\mathrm n} }{\mathrm d\xi^2}+ q^2 e_{\mathrm n} = \varepsilon \frac{\sin{(2k\xi)}}{2k}.
\label{eq.97}
\end{equation}
We recall that the Debye wave length $\lambda_{\mathrm D}$ was defined in Eq.~(\ref{eq.48}). We used as an abbreviation the complex wave vector $q =
i/\lambda_{\mathrm D}$, and further introduced the forcing amplitude as
\begin{equation}
\varepsilon  = \sum_j \frac{1}{\lambda^2_j} \frac{\Omega_j}{V}(b_1^2 - b_2^2),
\label{eq.98}
\end{equation}
which vanishes for a circularly polarized wave with $b_1=b_2$, but is nonzero
otherwise. The solution of Eq.~(\ref{eq.97}) can be obtained by using the Green's
function method and Fourier transformation. 
The convolution integral of the
forcing term with the Green's function then yields the solution depending
upon $\xi$ in the form of another convolution integral, which can be calculated
analytically with the (in $q$ symmetric) result
\begin{equation}
e_{\mathrm n}(\xi) =\frac{\varepsilon}{4kq}\left( \frac{ \sin{(2k \xi)} + \sin{(q \xi)}}{2k+q}
- \frac{\sin{(2k \xi)} - \sin{(q \xi)}}{2k-q} \right).
\label{eq.99}
\end{equation}

Eq.~(\ref{eq.99}) solves the original Eq.~(\ref{eq.97}) that can easily be
shown by straightforward differentiation. We recall that by definition the
square of the wave vector $q=q(V)$ is given by the right-hand side of the
electrostatic dispersion relation (\ref{eq.71}). Therefore, $q$ is a real
number for an appropriate choice of $V$. The solution is then related
naturally with the electrostatic eigenmodes, i.e. the sound, ion acoustic,
and Langmuir waves, which we already discussed in a previous section. The
overall solution (\ref{eq.99}) apparently describes forced compressive
(charge) waves, occurring as electrostatic eigenmodes, and a superposed
electric wave at the second harmonic of the transverse Alfv\'en pump wave,
the anisotropy (due to its elliptic polarization) of which determines the
amplitude of these driven longitudinal electric field oscillations.
L'H\^opital's rule allows one to determine the behavior in the resonant
cases. For the resonances ($q\rightarrow \pm 2k$), we find
\begin{equation}
e_{\mathrm n}(\xi) = - \frac{\varepsilon}{2q^2}\left[\xi
\cos(q\xi)-\frac{\sin(q\xi)}{q}\right],
\end{equation}
which corresponds to the amplitude of the compressible oscillation growing or
decaying with $\xi$, i.e. an instability in space.

\section{Discussion and conclusions}

Starting from the multi-fluid equations of a warm plasma, we have derived and
investigated the coupled wave equations for the particles' gyromotions about
the mean field and for the transverse magnetic field and longitudinal
electric field. It is a natural outcome of the electromotive forces arising
from compressible Alfv\'en-cyclotron waves and can be derived from a
potential that is just the kinetic energy associated with the gyromotion in
the electromagnetic wave. Electric waves are thus excited, which can react
back on the pump wave by nonlinear effects through terms in its own wave
equation that contains the electric field explicitly. Known limiting cases
are reproduced, such as the standard linear electric waves like the
ion-acoustic or Langmuir waves of course, and for the transverse magnetic
field the usual two branches of Alfv\'en-cyclotron and magnetosonic-whistler
waves in case of a two-component electron-proton plasma, or many similar
related branches in the case of a multi-ion plasma. The main result of this
paper is the closed set of second-order wave equations (\ref{eq.76}),
(\ref{eq.79a}), and (\ref{eq.79b}), from solutions of which the transverse
electric field and charge densities of each species can be derived as
auxiliary quantities. To study these wave equations in more detail and to
find their nonlinear solutions is left as a future task, which will
presumably require a numerical treatment.

The structure of our equations already permits to derive some qualitative
conclusions and to treat some simple applications (like the effect of
elliptical polarization) analytically. Further study is certainly required to
corroborate them quantitatively. Apparently, the weakly compressible
large-amplitude Alfv\'en-cyclotron waves can drive electric fluctuations,
essentially of the ion-acoustic type, along the mean field, and thus will
naturally produce an electric field that can accelerate particles and will
lead to heating via Landau damping in a kinetic Vlasov description. By
excitation of acoustic waves, the amplitude of the driver wave will be
diminished until a dynamic wave--wave equilibrium is reached. Similar
processes are clearly found in the direct numerical simulations
\citep{araneda08, araneda09, valentini08,valentini09}. The
third-order coupling terms in Eqs.~(\ref{eq.76}) and (\ref{eq.79b})
correspond to such three-wave processes in Fourier space, and therefore will
lead to cascading of spectral energy and broadening of the original spectrum
of the pump wave, which need not be monochromatic. This way a new path towards
micro- and macro-turbulence could be opened, and a non-MHD cascade is
rendered possible by these compressive Alfv\'en-cyclotron--acoustic wave
interactions.

\begin{acknowledgment}
D.~V. appreciates financial support by the International Max Planck Research
School (IMPRS) on Physical Processes in the Solar System and Beyond.
\end{acknowledgment}

\bibliographystyle{jpp}
\bibliography{alfven_cyclotron_waves_arxiv}

\end{document}